\documentclass{article}

\PassOptionsToPackage{numbers, compress}{natbib}

     \usepackage[final]{neurips_2023_ml4ps}

\usepackage[utf8]{inputenc} 
\usepackage[T1]{fontenc}    
\usepackage{hyperref}       
\usepackage{url}            
\usepackage{booktabs}       
\usepackage{amsfonts}       
\usepackage{nicefrac}       
\usepackage{microtype}      
\usepackage{xcolor}         
\usepackage{natbib}

\usepackage{amsmath}       

\usepackage{graphicx}
\usepackage{fontawesome}    

\title{Neural ODEs as a discovery tool to characterize the structure of the hot galactic wind of M82}

\author{%
  Dustin Nguyen\thanks{Corresponding Author}\\
  The Ohio State University\\
  \texttt{dnguyen.phys@gmail.com}\\ 
  \And 
  Yuan-Sen Ting\\
  Australian National University\\ 
  The Ohio State University \\ 
  \texttt{yuan-sen.ting@anu.edu.au} \\
  \And
  Todd A.~Thompson\\ 
  The Ohio State University\\ 
  \texttt{thompson.1847@osu.edu}\\
  \And
  Sebastian Lopez\\
  The Ohio State University\\
  \texttt{lopez.764@osu.edu}
  \And
  Laura A. Lopez\\
  The Ohio State University\\
  \texttt{lopez.513@osu.edu}
}

\begin{document}

\maketitle

\begin{abstract}
  Dynamic astrophysical phenomena are predominantly described by differential equations, yet our understanding of these systems is constrained by our incomplete grasp of non-linear physics and scarcity of comprehensive datasets. As such, advancing techniques in solving non-linear inverse problems becomes pivotal to addressing numerous outstanding questions in the field. In particular, modeling hot galactic winds is difficult because of unknown structure for various physical terms, and the lack of \textit{any} kinematic observational data. Additionally, the flow equations contain singularities that lead to numerical instability, making parameter sweeps non-trivial. We leverage differentiable programming, which enables neural networks to be embedded as individual terms within the governing coupled ordinary differential equations (ODEs), and show that this method can adeptly learn hidden physics. We robustly discern the structure of a mass-loading function which captures the physical effects of cloud destruction and entrainment into the hot superwind. Within a supervised learning framework, we formulate our loss function anchored on the astrophysical entropy ($K \propto P/\rho^{5/3}$). Our results demonstrate the efficacy of this approach, even in the absence of kinematic data $v$. We then apply these models to real Chandra X-Ray observations of starburst galaxy M82, providing the first systematic description of mass-loading within the superwind. This work further highlights neural ODEs as a useful discovery tool with mechanistic interpretability in non-linear inverse problems. We make our code public at this GitHub repository (\href{https://github.com/dustindnguyen/2023_NeurIPS_NeuralODEs_M82}{\texttt{github.com/dustindnguyen/2023\_NeurIPS\_NeuralODEs\_M82}} \href{https://github.com/dustindnguyen/2023_NeurIPS_NeuralODEs_M82}{\faGithub}).

\end{abstract}

\section{Introduction}
In the field of physics, understanding is often equated with having definitive differential equations that describe dynamic variables \citep{Meiss2007}. Yet, when it comes to modeling real-world systems, the true underlying physics is not always clear. This is especially the case for modern galactic wind models, where non-linear phenomena are difficult to describe analytically. Recent wind-cloud interaction simulations \citep{Gronke2018,Gronke2020,Gronke2022} indicate efficient non-linear multi-phase mixing can overcome the so-called cloud-crushing problem \citep{Klein1994} posed by observations of fast cool outflows from nearby starburst galaxies \citep{Lynds1963}. Subsequently, there has been an explosion in activity focused on running suites of high-resolution 3D time-dependent hydrodynamic simulations to extract useful scaling laws relevant to cloud survival and the interface of radiative turbulent mixing \citep{Schneider2020,Fielding2020,Abruzzo2022,Tan2023b,Tan2023}. These intuition-based and simulation-based analytic models offer valuable insights. However, considering observations of the cosmos already encodes the necessary information, a shift towards data-driven modeling, set apart from the customary data-testing approach in astronomy, combined with traditional physical models might provide fresh insights into the intrinsic structure of galactic winds.

Deep neural networks have been highlighted as powerful tools for approximating unknown functions and operators \citep{LeCun2015, Goodfellow2016, HORNIK1990, Lu2021}. A noteworthy application of these networks is their role in the numerical discretization of ODEs/PDEs, leveraging the adjoint sensitivity method \citep{Chen2019}. Physics-Informed Neural Networks (PINNs) emphasize the integration of differential equations equations into their loss functions and employ the adjoint sensitivity method for effective gradient retrieval \citep{Raissi2019}. Expanding on these ideas, the landscape has further matured to introduce the Universal Differential Equation (UDE) paradigm \citep{rackauckas2020universal}. Unlike PINNs, which primarily focus on embedding the structure of differential equations into the neural network's loss function, UDEs pioneer the fusion of neural networks directly into the differential equations themselves, significantly amplifying their approximation prowess. Once integrated, these networks inherently respect conservation laws and shape the system's dynamics sequentially, ensuring that each contribution can be meaningfully and mechanistically interpreted. Notably, while PINNs are unique in their solution approach, UDEs can be seamlessly solved using standard numerical methods (e.g., Runge-Kutta 4 [RK4]), bridging the gap between traditional differential equation solvers and deep learning.

The focus of this paper revolves around these UDEs, which in this paper, we refer to as simply "neural ODEs". The incorporation of neural networks into ODEs/PDEs through differentiable programming is an emerging research area with broad application in the physical and biological sciences \citep{Vortmeyer-kley2021, Gelbrecht2021, Keith2021, Fronk2023, Stepanmiants2023, Yin2023, Santana2023,Nguyen2023_icml}. In the realm of galactic wind simulations, 3D time-dependent models offer intricate insights into non-linear behaviors. However, 1D steady-state models have demonstrated their efficacy in accurately capturing the statistical global properties of these complex 3D simulations. Such 1D models provide a streamlined approach, effectively summarizing the essential characteristics and trends of galactic winds, and form the foundation for our exploration using neural networks. In this work we use neural ODEs to model real observations of a galaxy and attempt to characterize the structure of mass-loading, which captures the effect of cool cloud destruction, and subsequent entrainment, into a hot supernovae-driven galactic wind. We focus on characterizing X-ray observations of the widely studied starburst prototype galaxy: Messier 82 (M82). We use the astrophysical entropy ($K/k_b = T/n^{2/3}$) as a feature-engineered (physical) variable within the loss function and penalize diverging solutions. We will show that regression on the this loss function allows us to infer a mass-loading model (with no prior knowledge) that makes predictions which better match the observed temperature and density profiles.

\section{Neural Galactic Wind Model}
The hydrodynamic equations for a steady-state hot flow moving in the $x$ direction are \citep{Cowie1981,Nguyen2021,Nguyen2023_icml}: 
\begin{equation}
\frac{( A \rho v )_x}{A} = \dot{\mu} ,  \quad  v  v_x = -  \frac{P_x}{\rho} - \nabla \Phi - \frac{\dot{\mu} v} { \rho} , \quad \mathrm{and} \quad  v \epsilon_x - \frac{v P \rho_x} {\rho^2}  = - \frac{n^2 \Lambda}{\rho} + \frac{\dot{\mu}}{\rho} \bigg( v^2 - \epsilon - \frac{P}{\rho} \bigg),
    \label{eq:eqs}
\end{equation}
where $v$, $\rho$, $P$, $\epsilon$, $\nabla \Phi$, $n$, $\Lambda$, $A$, $\dot{\mu}$, and subscript $_x$ are the bulk velocity, density, pressure, specific internal energy, gravity, number density, cooling rate, surface area expansion rate, volumetric mass-loading rate, and first-order spatial derivative, respectively. Mass-loading term $\dot{\mu}$ captures the global effects of cool cloud destruction and incorporation into the hot phase \citep{Cowie1981,Schneider2020,Nguyen2021,Fielding2022}. In this work, it is assumed that the entrained cool material contains negligible velocity and temperature, which is valid in the limit that the hot galactic wind contains most of the thermal and kinetic energy \citep{Cowie1981}. We take $\nabla \Phi = \sigma ^2 / x$, where $\sigma = 200 \, \mathrm{km\,s^{-1}}$, polynomial fit \citep{Schneider2015} to the radiative cooling curve, and $A$ is defined below separately for the mock test and comparison to real data. When substituted into each other, each of the derivatives $v_x$, $\rho_x$, and $P_x$ are proportional to $(\mathcal{M}^2-1)^{-1}$ \citep{LamersCasinelli1999} and thus all contain a singularity at the sonic point (i.e., numerically diverges at $\mathcal{M}=1$). The dimensions of each simulation is 0.37\,kpc$\leq x \leq$2.65\,kpc and number of steps is $n_x = 500$. We do not use all $500$ points during optimization in both the mock test and comparison to real data, which only has 44 resolved data points. As there is currently no kinematic measurements, we use the classic galactic wind model by \citet{Chevalier1985} to guess an initial velocity of $v_\mathrm{hot} \sim 1835\,\mathrm{km\,s^{-1}}$ after leaving the starburst volume $R=0.3\,$kpc. The initial conditions are then $n_0 = 0.843 \, \mathrm{cm^{-3}}$, $T_0 = 0.615 \times 10^7\,\mathrm{K}$. In the mock test we take $v_0 = v_\mathrm{hot}$, and then for the Chandra data, we consider two initial velocities $v_{0,a} = v_\mathrm{hot}$ and $v_{0,b} = v_\mathrm{hot}/2$.

In summary, our method is:

\begin{enumerate}
    \item Solve an initial value problem (Eqs.\,~\ref{eq:eqs}) by RK4 integration of $v_0,\ \rho_0 , \ P_0$ from $x_0$ to $x_f$. 
    \item Calculate loss (Eqs.~\ref{eq:loss_MSE} and \ref{eq:loss_penalty}) between the data and integrated solutions at points of $n_{x,\mathrm{data}}$. 
    \item Backpropagate through automatic-differentiated ODE solver to get gradients.
    \item Update weights of individual neural network variable $\dot{\mu}$.
    \item Iterate for 150 epochs of \texttt{ADAM} optimizaton and then up to 150 epochs of \texttt{BFGS} optimization.
\end{enumerate}

We represent the volumetric mass-loading rate, $\dot{\mu}$, with a multi-layer perceptron neural network comprised of 3 hidden layers. The input into $\dot{\mu}$ is a single position, and the full range of positions is sampled by forward integration of Eqs.~\ref{eq:eqs} using 5th order RK4 \citep{TSITOURAS2011}. The bulk velocity $v$ cannot be used in the optimization problem because there is not yet any kinematic data available for M82's X-ray emitting wind. We calculate the loss function using a feature-engineered quantity, the astrophysical entropy $K/k_b = T/n^{2/3} \propto P/\rho^{5/3}$. The loss function is a weighted Mean-Square-Error (MSE): 

\begin{equation}
    \mathcal{L} = \sum_i^{n_{x,\mathrm{data}}} \bigg[ \mathcal{W}_i \times \bigg (K_i - \hat{K}_i \bigg)^2 \bigg]. 
    \label{eq:loss_MSE}
\end{equation}

where $K$ is the data, and $\hat{K}$ are the solutions of the integrated ODEs. The weights $\mathcal{W}$ linearly scale the MSE as a function of $n_x$, where $\mathcal{W}_0=1$ and $ \mathcal{W}_{n_{x,\mathrm{data}}} \ll 1$, $n_{x,\mathrm{data}}=44$ is the number of data points. This scaling increases sensitivity to early solutions, which is important for non-linear problems. We do not include division by $n_{x,\mathrm{data}}$ or $k_b^2$ because it does not impact training. If the flow at any instance reaches the sonic point, $\mathcal{M} =1$, the equations become numerically unstable \citep{LamersCasinelli1999}, hindering the optimization process. We prevent this by introducing a penalty term \citep{Nguyen2023_icml} that activates between $1 \leq  \hat{\mathcal{M}} \leq \mathcal{M}_\mathrm{penalty} $, where we take $\mathcal{M}_\mathrm{penalty} = 1.5$. In this region, the loss is artificially increased per optimization step by:
\begin{equation}
    \mathcal{L} = \mathcal{L}_\mathrm{previous} \times \omega \sum_i ^{n_{x,\mathrm{data}}} \bigg[ 1 - \bigg(1 - \hat{\mathcal{M}} \bigg)^2 \bigg] \quad \quad (1 \leq \hat{\mathcal{M}} \leq \mathcal{M}_\mathrm{penalty}),
    \label{eq:loss_penalty}
\end{equation}
where $\omega$ is a constant. We minimize the loss function using two optimization algorithms, which has been shown to be required for convergence in other neural ODE studies \citep{Rackauckas2021,Nguyen2023_icml}. We solve the equations, automatic-differentiate the system, and calculate gradients entirely within the \texttt{Julia} SciML ecosystem. 

\begin{figure}
    \centering
    \includegraphics[width=\textwidth]{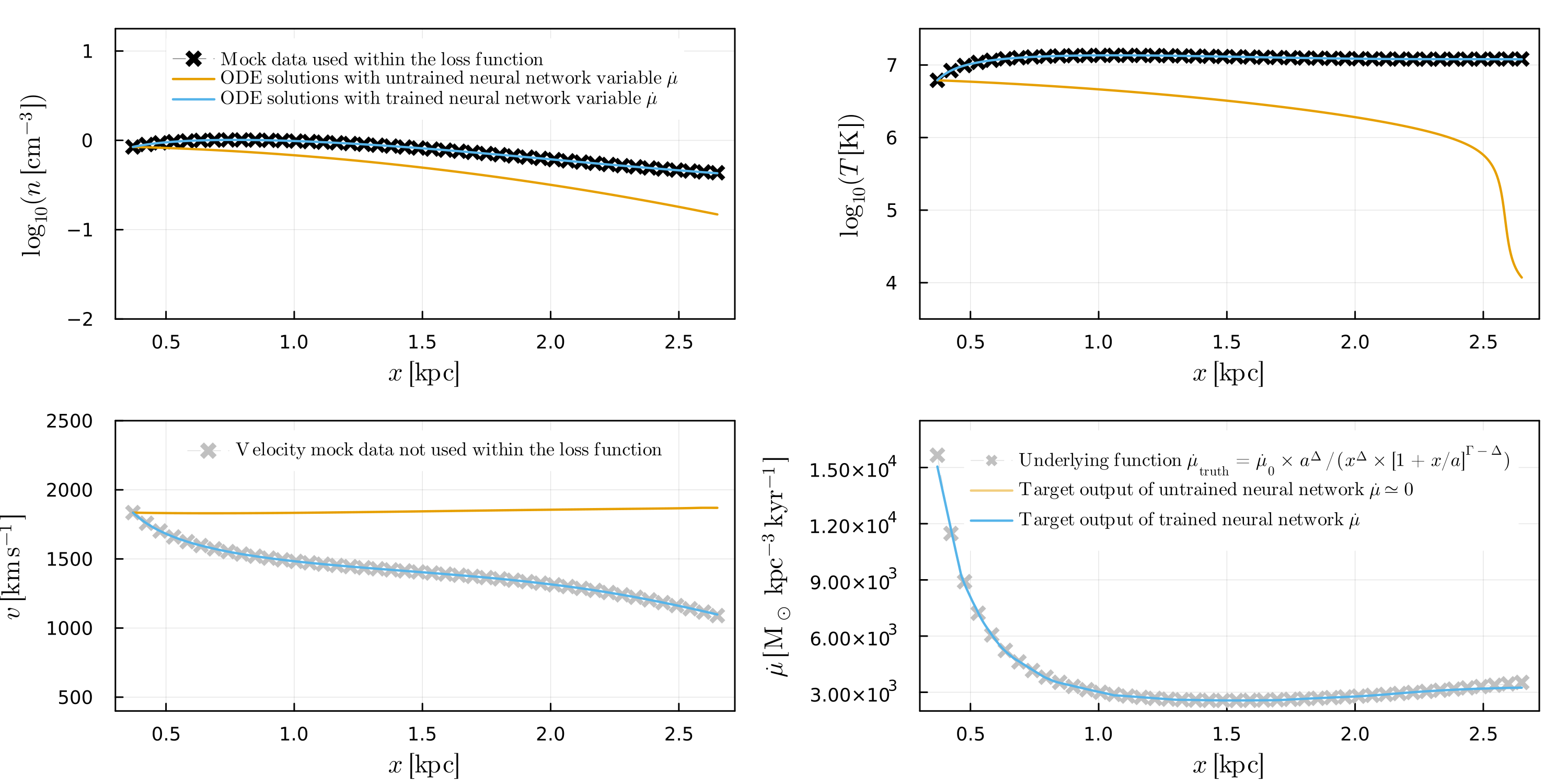}
    \caption{Density, temperature, and velocity profiles for neural ODE solutions and the target output of neural network variable $\dot{\mu}$ (bottom right panel). The x's are the mock data, of which, only the density and temperature is used in the loss function. Despite not using the kinematic mock data, the correct kinematics is predicted and the underlying mass-loading function was learned after training.} 
    \label{fig:mock}
\end{figure}

\section{Results}

\textbf{Mock Test: }We test the framework to learn a mass-loading function described by
\begin{equation}
\dot{\mu}_\mathrm{truth} = \dot{\mu}_0 \times a^\Delta / (x^\Delta \times (1 + x /a )^{\Gamma-\Delta})
\label{eq:mudot_test}
\end{equation}
where $\dot{\mu}_0 = 10 \, \mathrm{M_\odot \, kyr^{-1} \, kpc^{-3}}$, $a$ = 1.5\,kpc, $\Delta = 4.0$, and $\Gamma = -4.0$. This equation scales as $x^{-\Delta}$ and transitions to $x^{-\Gamma}$ at approximately distance $a$, describing intense cloud entrainment after the wind leaves the host galaxy \citep{Nguyen2021}. We take the flow geometry to be a flared-cylinder $A = A_0 \, (1+ (x/\eta)^2)$, which characterizes a cylindrically-collimated flow that later undergoes spherical expansion \citep{Kopp1976,Everett2008,Nguyen2022}. Substituting  Eq.~\ref{eq:mudot_test} into \ref{eq:eqs}, we calculate the mock-data (i.e., $\dot{\mu} \rightarrow \dot{\mu}_\mathrm{truth})$. We then ``forget" $\dot{\mu}_\mathrm{truth}$ and replace $\dot{\mu}$ with a neural network and initialize it such that the output across all $x$ is approximately 0 (using \texttt{Flux.jl}'s default \texttt{Glorot} initialization). The first (i.e., untrained) neural ODE solution will be that of a wind with approximately zero mass-loading. The mock data is calculated with $n_x=500$ steps, however, we utilize only $n_{x,\mathrm{data}}=44$ linearly spaced samples of the mock dataset during optimization to mimic the resolution of the Chandra X-ray data we later focus on. In Fig.~\ref{fig:mock} we plot the neural ODE solutions. Despite not using the kinematic mock data $v$ we still predict the correct kinematic, and thermodynamic, solutions and learn the underlying mass-loading function $\dot{\mu}_\mathrm{truth}$.

\begin{figure}
    \centering
    \includegraphics[width=\textwidth]{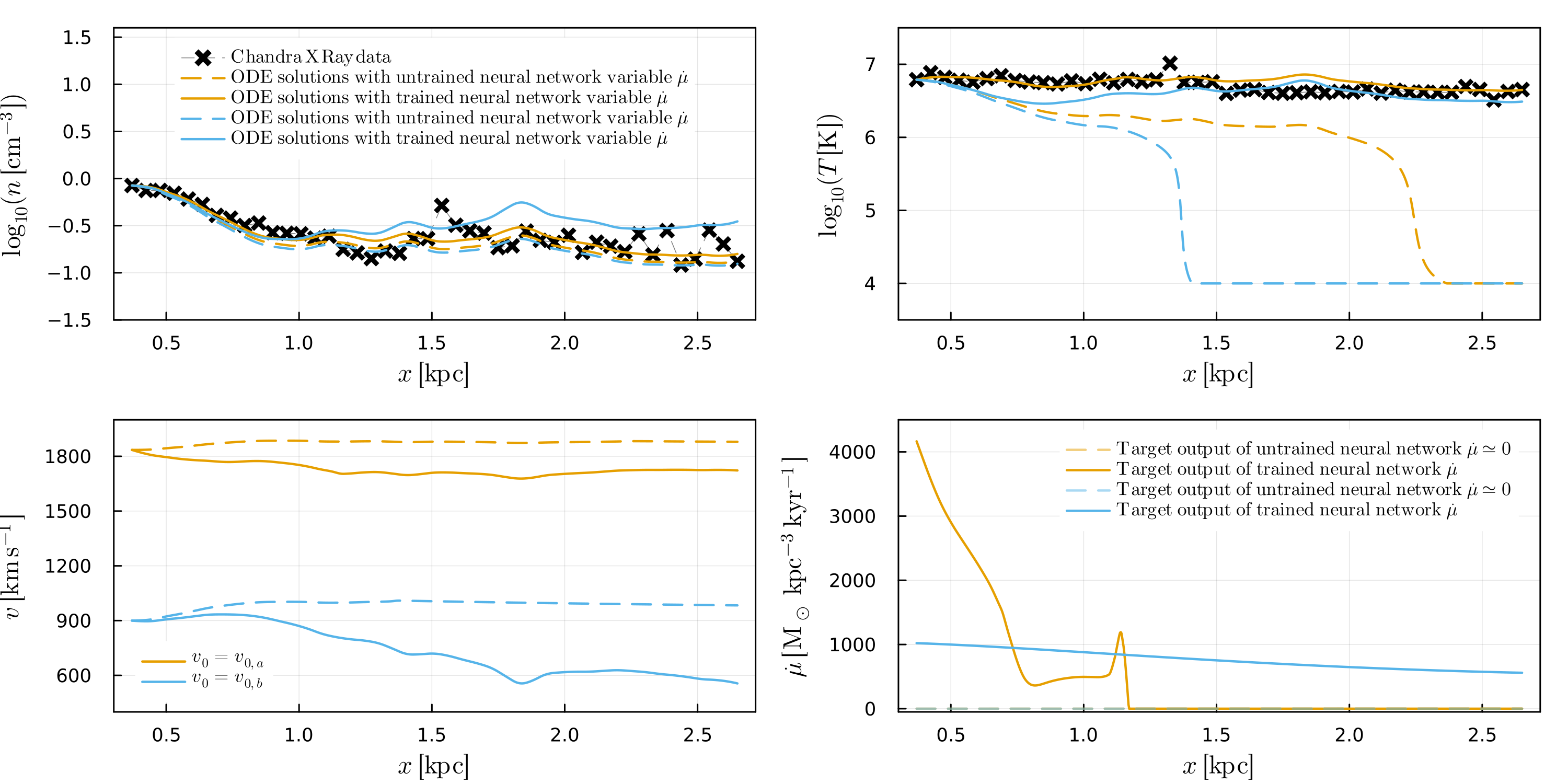}
    \caption{Density, temperature, and velocity profiles for neural ODE solutions and the target output of neural network variable $\dot{\mu}$ (bottom right panel) for two different initial velocities (yellow and blue lines). The x's are the Chandra X-ray data which exists only for temperature and density. After training, the learned mass-loading function leads to a better fit to the data, as the thermalization of kinetic energy prevents rapid bulk cooling that would otherwise cool the flow to $10^4\,$K.}
    \label{fig:results1}
\end{figure}

\textbf{Application towards Chandra X-Ray Data: }We now consider applications of neural ODEs towards Chandra X-ray observations of the northern wind of galaxy M82 (see \citep{Lopez2020}). There are 44 data points for $n$, $T$, and $A$. For illustrative purposes, we will consider two initial velocities: $v_{0,a} \sim v_\mathrm{hot}$ and $v_{0,b} \sim v_\mathrm{hot}/2$. We plot the neural ODE solutions for both scenarios in Fig.~\ref{fig:results1}. The additional heating from the learned mass-loading prevents rapid bulk cooling leads to better matched temperature profiles. For $v_0=v_{0,a}$, the target output of neural network variable $\dot{\mu}$ (yellow line) reveals non-trivial structure, sharply truncating at $x\sim 1.1\,$kpc. The mass-loading rate and initial wind mass-outflow rate is calculated as $\dot{M}_\mathrm{load} = \int dx \, A \, \dot{\mu} $ and $\dot{M}_\mathrm{wind} = A_0 \rho_0 v_0$. We find $\dot{M}_\mathrm{load} / \dot{M}_\mathrm{wind} = 0.09$ and 0.59 for initial velocities of $v_0 = v_{0,a}$ and $v_{0,b}$. In the latter case, the mass-loading rate is roughly half the initial wind outflow, suggesting a sharp metal abundance gradient as seen in \citep{Lopez2020}. 

\section{Conclusion}
In this work we use neural ODEs to explain X-ray observations of outflows from starburst galaxy M82. Rather than parameter estimation on an assumed function, here the data formulates mass-loading term $\dot{\mu}$ without any prior knowledge its structure. We start with approximately zero mass-loading, showcasing the flexibility of the model to discover rich structure (orange line, bottom right panel Fig.~\ref{fig:results1}) as an ab initio modeling process. This work highlights the exceptional utility of neural networks as universal function approximators for non-linear inverse problems. 

New measurements by future X-ray space missions such as the recently launched XRISM satellite \citep{XRISM2020} and LEM concept \citep{LEM2022}, will provide the first kinematics measurements of the hot $10^7\,$K gas. Our study indicates the learned mass-loading factors are sensitive to the assumed initial hot gas velocity. These forthcoming data can easily be integrated into our methods to better understand cool cloud entrainment, ultimately shedding light on important processes in launching multi-phase winds. 

\textbf{Comparison to previous work: } This paper employs neural networks as universal function approximators for individual terms in a steady-state galactic wind model, akin to the approach of \citep{Nguyen2023_icml}. The salient distinctions in our work include: (1) our flow equations encompass critical physical processes such as radiative cooling and gravity; (2) our optimization with mock data operates at roughly a tenth of the spatial resolution in \citep{Nguyen2023_icml}, $n_{x,\mathrm{data}}$, mirroring the resolution observed in real data; and (3) our model omits one of the three dynamical variables, $v$, reflecting the current absence of velocity data in X-ray observations.

\textbf{Neural ODE comparison to Fourier series: }In the context of a 1D steady-state problem like our galactic wind model, alternative universal function approximators, such as the Fourier series, may seem viable. However, in future direct comparisons with 2D X-ray surface brightness images of galaxies, more intricate models—2D or 3D spatial hydrodynamics—are essential to capture finer structure. When considering time, these evolve into 3D and 4D models, respectively. Fourier series becomes less effective in these multidimensional contexts due to its inefficiency in scaling beyond 1D. Furthermore, it's imperative to note that while our study employs neural networks strictly as universal function approximators, there's a burgeoning effort in leveraging them for symbolic regression within ODE/PDE systems (see \texttt{PySR} and \texttt{SymbolicRegression.jl}), as evidenced by works such as \citep{rackauckas2020universal,cranmerDiscovering2020,cranmerInterpretableMachineLearning2023,Santana2023}. Employing symbolic regression on these trained neural networks promises to yield symbolic, interpretable expressions that elucidate the underlying physics being learned. While deriving such symbolic interpretations was not the primary objective of our study, the foundational step of training the neural network within the confines of the physical system was the central focus of this work. We also note that key integrated quantities, such as the total mass being deposited into the hot wind, can still be inferred without symbolic representations of the trained neural network.

\section*{Appendix}
\textbf{Data: } Using 534 ks of archival Chandra X-ray Observatory data, we constrain the temperature and density gradient along the outflow spamming $\pm 2.6\,$kpc. Spectra was extracted using CIAO for 101 rectangular regions along the outflow, each of size $3''\times1'$. To constrain the temperature and density, each region's spectra was modeled in XSPEC using $\texttt{const}*\texttt{phabs}*\texttt{phabs}(\texttt{powerlaw}+\texttt{vapec})$, where the abundances were frozen based on \citep{Lopez2020}. We focus on the northern outflow of M82. The southern side is largely asymmetric due to tidal interaction with M81, requiring a treatment that is beyond the scope of this work.

\section*{Acknowledgements}
DN acknowledges funding from NASA 21-ASTRO21-0174. Y.S.T. acknowledges financial support from the Australian Research Council through DECRA Fellowship DE220101520.





\bibliographystyle{unsrtnat}
\bibliography{dustin.bib}

\end{document}